\documentclass{article}
\usepackage{ijcai21}

\usepackage{times}
\usepackage{soul}
\usepackage{url}
\usepackage[hidelinks]{hyperref}
\usepackage[utf8]{inputenc}
\usepackage[small]{caption}
\usepackage{graphicx}
\usepackage{amsmath}
\usepackage{amsthm}
\usepackage{booktabs}
\usepackage{algorithmic}
\usepackage[ruled,linesnumbered,noend]{algorithm2e}
\newtheorem{theorem}{Theorem}

\newtheorem{corollary}{Corollary}

\usepackage{helvet}  
\usepackage{courier}  

\usepackage{multirow}
\usepackage{subcaption}
\usepackage{xcolor}

\newtheorem{definition}{Definition}

\usepackage{xspace}
\newcommand{\cppp}{\textsc {cppp}\xspace}
\newcommand{\mafs}{\textsc {mafs}\xspace}

\newcommand{\mastrips}{\textsc {ma-strips}\xspace}
\newcommand{\dpp}{\textsc {DPP}\xspace}

\newcommand{\commentout}[1]{}

\usepackage{array}
\newcolumntype{L}{>{\centering\arraybackslash}m{3cm}}

\commentout{
 
\declaretheoremstyle[%
  spaceabove=0pt,%
  spacebelow=0pt,%
  headfont=\normalfont\itshape,%
  postheadspace=1em,%
  qed=\qedsymbol%
]{mystyle} 

}
\usepackage{adjustbox}

\theoremstyle{remark}

\newcolumntype{R}[2]{%
    >{\adjustbox{angle=#1,lap=\width-(#2)}\bgroup}%
    l%
    <{\egroup}%
}

\newcommand{\citet}[1]{\citeauthor{#1}~\shortcite{#1}}
\newcommand{\citep}[1]{\cite{#1}}

\title{Partial Disclosure of Private Dependencies in Privacy Preserving Planning}

\author{
Rotem Lev Lehman$^1$\and
Guy Shani$^1$\And
Roni Stern$^{1,2}$
\affiliations
$^1$Software and Information Systems Engineering, Ben Gurion University of the Negev, Be'er Sheva, Israel\\
$^2$Palo Alto Research Center, Palo Alto, CA, USA
\emails
levlerot@post.bgu.ac.il,
shanigu@bgu.ac.il,
sternron@post.bgu.ac.il
}

\begin{document}

\maketitle

\begin{abstract}

In collaborative privacy preserving planning (\cppp), a group of agents jointly creates a plan to achieve a set of goals while preserving each others' privacy. During planning, agents often reveal the private dependencies between their public actions to other agents, that is, which public action facilitates the preconditions of another public action.
Previous work in \cppp does not limit the disclosure of such dependencies. In this paper, we explicitly limit the amount of disclosed dependencies, allowing agents to publish only a part of their private dependencies. 
We investigate different strategies for deciding which dependencies to publish, and how they affect the ability to find solutions. 
We evaluate the ability of two solvers --- distribute forward search and centralized planning based on a single-agent projection -- to produce plans under this constraint.
Experiments over standard \cppp domains show that the proposed dependency-sharing strategies enable generating plans while sharing only a small fraction of all private dependencies.
\end{abstract}






\section{Introduction}

Designing autonomous agents that act collaboratively to achieve a common set of goals has been a major goal for Artificial Intelligence research for many years. 
\emph{Collaborative Privacy-Preserving Planning} (\cppp) is a multi-agent planning task in which agents need to achieve a common set of goals without revealing certain private information~\citep{brafman2008one}. In particular, in \cppp\ an individual agent may have a set of private facts and actions that it does not share with the other agents. \cppp\ has important motivating examples, such as planning for organizations that outsource some of their tasks, 
 and has recently received considerable attention from the academic community~\cite{nissim2014distributed,Brafman15,Torreno,tozicka2017theLimits,maliah2018action,caspari2020aFramework}.

There are two common approaches to \cppp: \emph{single search} and \emph{two-level search}.  
Single search solvers run a joint forward search \cite{nissim2014distributed,MADLA} where agents that apply public actions send the resulting state to other agents that continue the forward search. 
Two-level search solvers create a public plan skeleton that is shared by all agents, and then each agent extends it locally with private actions~\cite{maliah2015privacy}.

In either case, the agents either indirectly or explicitly publish \emph{dependencies} between the public actions. 
For example, in a multiagent Mars Rover scenario, the need to take a camera from a base station before sending an image, incurs a {\em private dependency} between these public actions. In both single search and a two-level approach some private dependencies are revealed during planning.
\citet{maliah2016projections} compute and publish all such private dependencies in the form of artificial facts. This allows the agents to jointly create and publish a \emph{projection} of the problem that contains all the public facts, public actions, and published artificial facts that capture private dependencies. 
Such a projection can be used to construct heuristics for single search \cppp algorithms such as \mafs. 
In two-level search solvers, this projection can be used to generate the public plan, defining the public plan to be a solution to the problem defined by the projection. 

In many cases, however, the agents can construct a plan requiring only a small portion of the private dependencies. It may be preferable to reveal only a part of the dependencies, intuitively reducing the amount of disclosed private information. 
This raises the challenge of how to choose which private dependencies to share and which not to. 
We suggest 4 different heuristic methods for deciding which dependencies should be published first.
We provide experiments on standard benchmarks from the \cppp\ literature. Our results show that in many domains using our heuristics allows computing a plan while publishing only a small portion of the private dependencies. 
We also analyze the makespan cost of the plans, showing that for some domains publishing fewer dependencies does not increase the plan cost significantly.

\section{Background}

An \mastrips\ problem~\citep{BrafmanD13} is a tuple $\langle P, \{A_i\}_{i=1}^k, I ,G \rangle$ where:
\begin{itemize}
	\item $k$ is the number of agents.
	\item $P$ is a finite set of primitive propositions (facts).
    \item $A_i$ is the set of actions agent $i$ can perform. 
	\item $I$ is the start state.
	\item $G$ is the goal condition.	
\end{itemize}

Each action $a=\langle pre(a), \mathit{eff}(a) \rangle$ is defined by its preconditions ($pre(a)$), and effects ($\mathit{eff}(a)$). Preconditions and effects are conjunctions of primitive propositions and literals, respectively. A state is a truth assignment over $P$.  $G$ is a conjunction of facts. $a(s)$ denotes the result of applying action $a$ to state $s$. A {\em plan} $\pi=(a_1,\ldots,a_k)$ is a solution to a planning task iff $G\subseteq a_k(\ldots(a_1(I)\ldots)$.

Privacy-preserving \mastrips\ extends \mastrips\ by defining sets of variables and actions as private, known only to a single agent. 
$private_i(P)$ and $private_i(A_i)$ denote the variables and actions, respectively, that are private to agent $i$. $public(P)$ is the set of public facts in $P$. $public_i(A_i)$, the complement of 
$private_i(A_i)$ w.r.t.~$A_i$, is the set of public actions of agent $i$. Some preconditions and effects of public actions may be private,
and the action obtained by removing these private elements is called its {\em public projection\/}, and it is known to other agents. When a public action is executed, all agents are aware of the execution, and view the public effects of the action. The goals can be public, but can also be private to a single agent.
An agent is aware only of its {\em local view} of the problem, that is, its private actions and facts, its public actions, the public facts, and the public projection of the actions of all other agents. That is, for public actions of other agents, the agent's local view contains only the public preconditions and effects of these actions.

\begin{figure}
    \centering
    \includegraphics[scale=0.3]{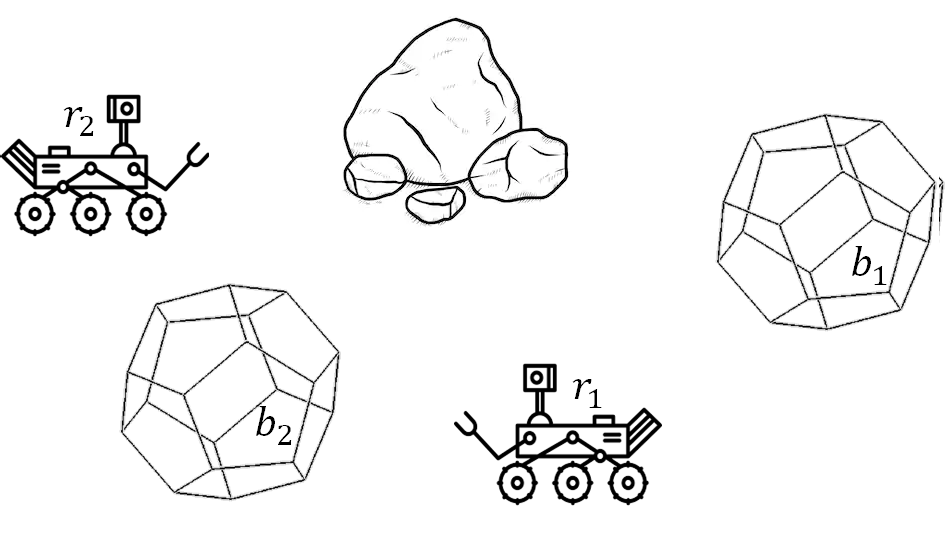}
    \caption{The rovers domain, with 2 base stations $b_1$ and $b_2$, and 2 rovers, $r_1$ and $r_2$,   collaborating to take measurements of a rock.}
    \label{fig:RoversDomain}
\end{figure}
	
In a Rovers example (Figure~\ref{fig:RoversDomain}), 2 Mars Rovers collaborate to explore Mars rocks. The Rovers need to perform sensor measurements on rocks, and, due to limited carriage capacity, can only carry 2 sensors at a time. Unused sensors are stored in base stations, and can be taken and returned to the base stations as needed. 
The public facts in this problem are the sensors located in bases, and the current condition of the target rock. The public actions are taking and returning sensors to the base stations, putting collected rock samples at the base stations, and performing various examination actions on rocks, such as taking an image, mining a mineral, or collecting a sample. The sensors held by a rover and its position are private, and the private actions are movement actions.

\subsection{Privacy Models}

An algorithm is privacy-preserving, if it provably does not ``reveal private information''. Much of the research on privacy-preserving planning considered revealing private information  only if the information is explicitly communicated to another agent. For example, if an agent  publishes during planning that it intends to drive rover $r_1$ from the base $base_1$ to the rock $rock_1$, then clearly the agent has revealed the existence of $r_1$, as well as an ability to achieve the private fact {\em (at $r_1$ $rock_1$)}, breaking the privacy constraint. However, if the agent only publishes that it can achieve a private fact with the obfuscated name $p$, then it is unclear what private information has been revealed. Thus, some privacy-preserving MA-STRIPS planners are built on {\em obfuscating} the private information they publish by applying some cryptographic tool~\citep{luis2014planMerging,Borrajo2015MAPR_CMAP}. 
	
	\citet{Brafman15} shows that the above form of privacy is weak, in the sense that there is no well-formed constraint on what other agents can {\em infer} from the information available to them. For example, if the public plan consists of agent $i$ taking a camera and the {\em take} action requires a rover to be present at the base that contains the camera, then all agents now know that $i$ controls at least one rover. Brafman considered a stronger form of privacy, where a fact or a specific value of a fact is {\em strongly private} if other agents cannot deduce its existence from the available information.
		
	By ``deducing the existence'' of a private fact, we mean that regardless of the reasoning power of the agents, they cannot infer the existence of a strongly private fact from the available information. The information available to an agent is (1) its local view of the problem, (2) the messages between the agents during planning, and (3) the sequence of public actions (of all agents) in the resulting plan. 
     A multi-agent planning algorithm is said to be {\em strongly privacy preserving} if the only information that agents can deduce, following the execution of the algorithm, is information that is implied by the public projection, their local view, and the public part of the solution~\citep{Brafman15}. 
	
	While appealing, achieving such a strong form of privacy may be difficult. 
    In fact, only two algorithms proven so far have this strict form of privacy: Secure \mafs\ ~\citep{Brafman15} and 
    two special variants of PSM~\citep{tozicka2017theLimits}. Even these algorithms preserve this form of privacy only under restrictive conditions. 
    For example, secure \mafs\ was shown to preserves strong privacy in a short list of specific domains (logistics, satellites, rovers), having unit action cost and when the heuristic function ignores the private state. 
    
    Thus, using more informed heuristics, such as the dependencies preserving projection heuristic~\citep{DPP} that we discuss later in the article, may violate the strong privacy of Secure MAFS. The current planners that preserve strong privacy are either inefficient or incomplete~\citep{tozicka2017theLimits,Leakage}. 
    Inefficiency in this context is that for any \mastrips\ problem, all public solutions to all local views must be computed before a solution is returned. \citet{tozicka2017theLimits} suggested considering cases where an algorithm maintains strong privacy for a class of problems.
  
Researchers have offered other definitions of privacy aside from the two extreme cases of weak and strong privacy. \citet{FwdBwd} suggest that an agent should only be aware of {\em neighboring} agents that modify a subset-private fact that the agent uses as precondition or effect. 
An algorithm preserves {\em agent privacy} if no agent can infer the existence of another agent with which it does not share subset-private facts.

Alternatively, \citet{DPP} suggest that agents may be aware of the types of objects that other agents manipulate, but not of their cardinality. In the rovers example, even though all agents may be aware that rovers carry tools between the base and the rocks, they should not be aware of the number of rovers, or the amount of tools in the rovers' storage. An algorithm preserves {\em cardinality privacy} if no agent can infer the number of private objects controlled by another agent. In our running example, no agent should know whether $agent_1$ controls 1 or 2 rovers.

Finally, instead of designing binary privacy criteria, one can consider more refined privacy metrics, that quantify the amount of leaked information \citep{vstolba2018quantifying}. For example, in distributed search \citep{maheswaran2006privacy} one often quantifies information loss using the entropy over the possible state space.
In that case, it may be possible for an application designer to sacrifice some privacy in favor of efficiency, such as the ability to scale up to larger problems.
The trade-off between privacy loss and efficiency is also studied in our paper, where privacy loss is quantified by the amount of revealed private dependencies and efficiency is measured by coverage and plan cost.

\subsection{Algorithmic Approaches}

We provide here a brief background on the two main approaches for \cppp: \emph{single search} and \emph{two-level search}. 
Single search algorithms for \cppp work by running a distributed forward search algorithm 
in which each agent searches for a plan using its own action space, and agents informs other agents of advancements in the search process by sending and receiving the states they have reached. \mafs\ \citep{nissim2014distributed} is a well-known example of a single-search algorithm for \cppp. In \mafs each agent runs a best-first forward search to reach the goal and maintains its own open list of states. In every iteration, each agent chooses a state in its open list to expand, generates all its children, and adds them to its open list (avoiding duplicates). Whenever an agent expands a state that was generated by applying a public action, it also broadcasts this state to all other agents. 
When an agent receives a broadcasted state, it adds that state to its open list. 
For example, in Figure~\ref{fig:RoversDomain}, when agent $1$ puts down the camera at base $b_2$, it broadcasts this state to all other agents. Agent $2$ can now use this state to pick up the camera and take a photo of a rock.
To preserve privacy, the private part of a state is obfuscated when broadcasting it by replacing the private facts with an index. Only the broadcasting agent knows how to map this index to the corresponding private facts. Once the goal is reached, the agent achieving the goal informs all others. 

Two-level search algorithms for \cppp work by computing a public plan, known as a coordination scheme \citep{nissim2014distributed,BrafmanD13,torreno2014fmap}, and then have each agent independently extend the public plan into a complete plan by adding private actions. 
In this extension each agent attempts to achieve the preconditions of its own public actions in the public plan. 
The Dependency Projection Planner (\dpp) \citep{maliah2018action} is an example of a two-level search algorithm. In \dpp, the agents compute together a single agent projection of the \cppp problem that captures the dependencies between public actions. 
This projection, referred to as the \emph{DP projection}, captures for each agent which public actions facilitate the execution of other public actions of that agent. 
In our running example, such a dependency exists, e.g., for agent $1$ between picking up a camera at base $b_1$ and taking a photo of the rock.
These dependencies are computed using limited regression from the precondition of a public action to the effects of other public actions. One can compute a public plan over the projection using a standard classical planner. The projection is incomplete, and hence the generated public plan may not be extended to a complete plan.

\section{Partial Disclosure of Private Dependencies}

We now present the main contribution of this paper --- reducing the amount of disclosed private dependencies and hence, the amount of disclosed information.
In this paper, we discuss this in the context of two state-of-the-art methods: \mafs~\cite{nissim2014distributed} and \dpp~\cite{maliah2018action}. 
We first describe our methods in the context of \dpp, and then discuss their application to \mafs .

\dpp works by generating a DP projection that is shared between all agents.  
For ease of exposition, we describe a brief and slightly modified version of the DP projection. We say that a public action $a$ {\em facilitates} the achievement of a private fact $f$, if (1) $f$ is an effect of $a$, or (2) there exists a sequence of private actions $a_1,...,a_k$ such that: $f$ is an effect of $a_k$,  each $a_i$ takes as precondition an effect of some $a_j$ s.t. $j<i$, and $a_1$ takes as precondition an effect of $a$.

\begin{definition}[Private Dependency]
An action $a$ is said to have a private dependency if it has a private fact $f$ as a precondition such that one of the following hold: (1) there is another public action $a'$ that facilitates achieving $f$,  (2) $f$ is either true in the start state or can be achieved from it by only applying private actions. 
\label{def:private-dependency}
\end{definition}

In the {\em rovers} example {\em take-image}$(rover_1,rock_1)$ 
has a private dependency because it has a private precondition  {\em holding}$(rover_1, camera_1)$
that the public action\\ {\em take}$(rover_1, camera_1, base_1)$ facilitates to achieve.

The agents jointly create a projection of the public problem, containing all the public facts, and a projected version of the public actions.
For each public action $a_i$ that requires a private precondition $f_j$, we create an artificial public fact $f^i_j$. The projected public version of $a$ requires $f^i_j$ as precondition. For each action $a'$ of the agent that facilitates the achievement of the private $f_j$, the agent publishes $f^i_j$ as an effect of $a'$, thus publishing the private dependency of $a$ on $a'$, although the way that $f_j$ is achieved remains obscured. 

For a public action $a_1$ of an agent $i$ we add to the projection a public artificial fact $f_{a_1}$, signifying that $a_1$ was executed. If $a_1$ facilitates the achievement of a private precondition of an action $a_2$ of $i$, then the projected $a_2$ will have $f_{a_1}$ as precondition. These artificial facts $f_a$ capture private dependencies between public actions.

In \dpp~\cite{maliah2016projections}, the agents compute and publish all their private dependencies. In this work, we limit the number of private dependencies of each agent is allowed to publish to $k$, where $k$ is a parameter. 
Technically, each agent publishes all the artificial preconditions of all public actions, as well as $k$ artificial effects of public actions. 
All public preconditions are published to avoid an over optimistic projection, where agents believe that they can execute public actions with no previous requirements.

In \mafs~\cite{nissim2014distributed}, each agent decides which dependencies it is willing to disclose, but it does not publish them. 
Instead, these dependencies are considered in \mafs when it generates states and when it computes heuristic values for a state. 
A state is only expanded if it does not reveal a dependency that is not disclosed, and it only considers published dependencies when computing heuristics for states. Further details on how \mafs considers the disclosed set of dependencies is given in section~\ref{scn:RevisedMAFS}.

\begin{figure}[ht]
    \centering
    \includegraphics[scale=0.4]{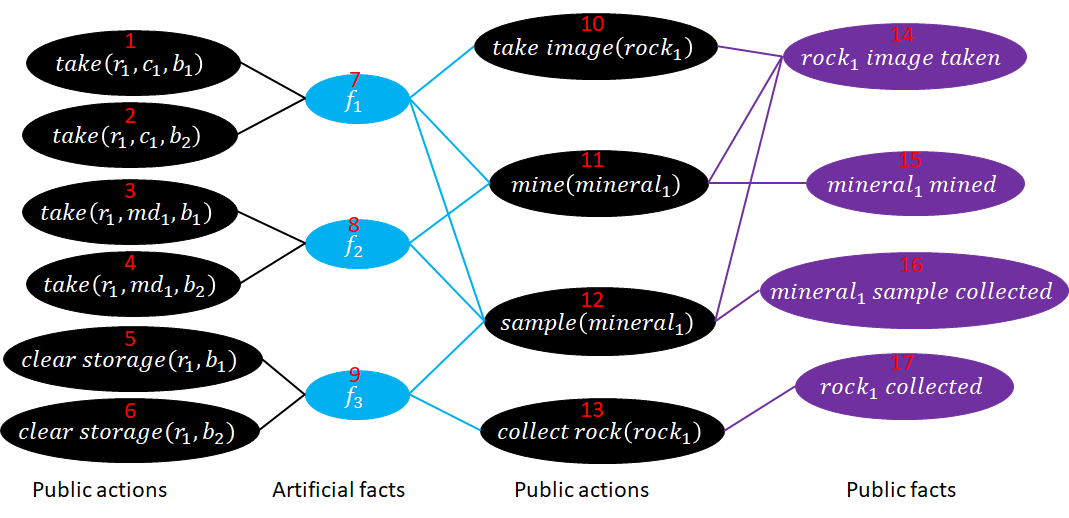}
    \caption{Local perspective of agent 1 private dependencies in the Rovers domain. Each node $n$ is marked with a unique identifier colored red. $r_1 = rover_1, c_1 = camera_1, b_i = base_i, md_1 = mineral\_detector_1$}
    \label{fig:dep}
\end{figure}

\subsection{Ranking Published Dependencies}
\label{scn:RankPublishedDep}

We now discuss how to choose which $k$ private dependencies to share. We suggest 4 different methods for selecting which dependencies to publish first. We take an iterative approach --- all agents publish one artificial effect of one public action at each iteration. Our methods assign a heuristic score to each dependency. The artificial effect with the highest score is published at each iteration. If the public projection cannot be solved, all agents recompute the heuristic scores, and publish a second artificial effect of a public action, and so forth. Hence, at each iteration an agent recomputes the scores, given the effects it has published thus far.

To better illustrate our methods, Figure~\ref{fig:dep} shows the local perspective of the private dependencies of agent 1 in the Rovers domain. Black nodes in the first and third columns represent public actions, purple nodes in the fourth column represent public facts, and blue nodes in the second column denote artificial facts that capture private dependencies between public actions. 
The public actions on the first column generate the artificial facts while the public actions on the third column require them as preconditions, and generate the public facts. Our methods always publish one of the black edges in the graph, between a public action and an artificial fact that it can generate.  The blue and purple edges (artificial preconditions and public effects) are always known.

Our first method, which we denote $m_1$, publishes artificial effects that are used as preconditions in as many public actions as possible. Initially the score of an artificial effect is set to the out-degree of the artificial fact. In the example in Figure~\ref{fig:dep}, we can publish either the effect of $n_1$, or the effect of $n_2$, representing taking the camera from either base, as it supplies a precondition for 3 public actions ($n_{10}$, $n_{11}$, $n_{12}$). Let us assume that the effect of $n_1$ was published first. 

We want to avoid choosing to publish the effect of $n_2$ at the next iteration, encouraging facilitating additional preconditions. We hence decrease from the score of an edge $\langle n_i, n_j \rangle$ by the number of incoming edges to the artificial fact $n_j$ that were already published. In our example, at the next step, all unpublished effects have the same ranking.

The second method, which we denote $m_2$, publishes an effect that  enables the achievement of as many public facts as possible, denoted by outgoing purple edges in Figure~\ref{fig:dep}. An effect enables the achievement of a public fact $f$ if it is a precondition to an action that achieves $f$. 
In $m_2$ we set the score of a black edge to the number of paths that lead from the edge to a public fact (purple node).
Again, at each iteration we decrease the score by number of incoming  edges that were published to avoid repeatedly choosing the same effect.
In our example all edges have the same score (3) at the first iteration.
If we again select at the first iteration the effect of $n_1$, at the second iteration we will select either $n_8$ or $n_9$.

The third method, denoted $m_3$, maximizes the amount of public actions that can be executed. That is, instead of publishing the artificial fact that provides a precondition for several actions, we publish the artificial fact that enables as many public actions as possible. Here, in the first iteration, publishing the effect of taking the camera from either base ($n_7$), or the effect of clearing the storage of the rover ($n_9$), both enable immediately one public action ($n_{10}$, and $n_{13}$, respectively). If we first select, again, $n_1$, at the next iteration, taking the mineral detector from either base station (producing $n_8$) or clearing the storage (producing $n_9$), both enable one additional action ($n_{11}$ and $n_{13}$ respectively), and are hence tied. 

In this method we also need to balance between enabling new public actions that have not been available given the already published effects, and enabling new ways to apply already possible actions, that may result in better plans. As such, for each action $a$ that is enabled by the published effect, we discount its score by $\frac{1}{c_a+1}$ where $c_a$ is the number of times that $a$ was enabled by previously published facts. Hence, at the first time that an action is enabled, it provides a score of 1, at the second time, a score of $\frac{1}{2}$, and so forth.

The last method, denoted $m_4$, is similar to $m_3$, but focuses not on the public actions, but on the public effects. That is, we publish an effect that enables achieving as many public facts (purple nodes) as possible. Again, we discount the score of a public fact by $\frac{1}{c_f+1}$ where $c_f$ is the number of times that public fact $f$ was enabled by previously published facts. In this method, again any incoming edge to $n_7$ and $n_9$ are tied in the first iteration. In the second iteration, however, if we again select $n_1$ first, taking the mineral detector (producing $n_8$) has a better score, because it enables the mineral drilling action ($n_{11}$), that has 2 public effects $n_{15}$, and $n_{14}$ that was already obtained, and hence a score of $1\frac{1}{2}$.


\subsection{The Implications of Adding Dependencies}

A DP projection $\Pi$ is a single-agent problem. 
Let $plans(\Pi)$, $opt(\Pi)$, and $states(\Pi)$ be the set of plans for $\Pi$, the cost of an optimal plan for $\Pi$, and the number of reachable states in $\Pi$, respectively.  
For two DP projections $\Pi$ and $\Pi'$ of the same problem, we say that $\Pi'$ \emph{subsumes} $\Pi$, denote by $\Pi\subseteq \Pi'$, iff the set of private dependencies shared by $\Pi$ is a subset of the set of private dependencies shared by $\Pi'$.

\begin{corollary}
For any two DP projections $\Pi$ and $\Pi'$ such that $\Pi \subseteq \Pi'$ it holds that: (1) $plans(\Pi) \subseteq plans(\Pi')$, 
(2) $opt(\Pi')\leq opt(\Pi)$, 
and (3) $states(\Pi) \leq states(\Pi')$.
\label{theorem:solvability}
\end{corollary}

This observation highlights the tradeoff we face when revealing more private dependencies. On the one hand, revealing more dependencies has the potential to solve more problems since more plans are reachable and find lower-cost solutions. On the other hand, DP projections created by revealing more dependencies are, in general, harder to solve, since its corresponding state space is larger. 
Corollary~\ref{theorem:solvability} is directly applicable
to all the heuristic methods described above for choosing which dependencies to share. For all methods ($m_1$, $m_2$, $m_3$, and $m_4$), increasing the number of dependencies that can be published ($k$) results in a DP projection that subsumes the DP projection created by publishing fewer dependencies. Indeed, we observe the aforementioned trade-off in our experimental results.

\subsection{Partial Disclosure of Dependencies in \mafs}
\label{scn:RevisedMAFS}

In this section, we provide more details on how we implemented partial disclosure of private dependencies in \mafs\ \citep{nissim2014distributed}. 

\begin{algorithm}[b!]
		\caption{MAFS for agent $r_i$ }
		\label{alg:MAFS}
		
		\SetKwBlock{MAFS}{MAFS($r_i$)}{end}
		\MAFS{
		    Initialize a set of dependencies $D$ that are not allowed to be revealed\label{line:initDependenciesNotAllowed}\\
			Insert $I$ into open list\\
			\While{solution not found}{ \label{line:final_verification} 
				\textbf{foreach} message $m$ call \textbf{process-message($m$)}\\
				$s\leftarrow$ \textbf{extract-min}(open list) \label{line:extract_min}\\
				\textbf{expand($s$)}\\
			}
		}
		
		\SetKwBlock{process}{process-message($m=\langle s,g_{j}(s),h_{j}(s)\rangle$)}{end}
		
		\label{alg:process-message}
		\process{
			
			\If{$s$ is not in open or closed list $\vee$ $g_{i}(s)>g_{j}(s)$  } { 
				add $s$ to open list and calculate $h_{i}(s)$ \label{line:local-h}\\
				$g_{i}(s)\leftarrow g_{j}(s)$\\
				$h_{i}(s)\leftarrow max(h_{i}(s), h_{j}(s))$\label{line:update-h}\\
			}
		}
		\SetKwBlock{expand}{expand($s$)}{end}
		\label{alg:expand}
		\expand{ 
			move $s$ to closed list\\
			\If{$s$ is a goal state}{
				broadcast $s, g_i(s), h_i(s)$ to all agents \label{line:broadcast1}\\
				\Return $s$ as the solution \label{line:return}
			}
			$a \leftarrow$ the action that generated $s$\\
			\If{$a$ is a public action}{
			    $a' \leftarrow$ the public action preceding $a$\\
			    \If{$\langle a',a \rangle \notin D$\label{line:testDependency}}{
			        send $s$ to all relevant agents
			    }
			    \Else{\Return~~~~~//~stop expansion of s}
			}
			apply $r_i$'s successor operators to $s$ \label{line:succ}\\
			\ForEach{successors $s'$}{\label{line:successors-forloop}
				compute $h_i(s')$\label{line:compute-h}\\
				\If{$s'$ is not in closed list or $h_i(s')$ has improved}{
					add $s'$ to open list \label{line:move-to-open-list}\\
				}
			}
		}
	\end{algorithm}

An agent running \mafs (Algorithm~\ref{alg:MAFS}) searches its own local space. Whenever agent $i$ executes a public action it broadcasts a message $m=\langle a_p, s_p, idx_1, ..., idx_n \rangle$ containing the executed public action $a_p$, the resulting public state $s_p$ to all other agents, with a private state index $idx_k$ for every agent $k$. Agent $i$ can now continue searching from that state. If, in the same search path, agent $i$ executes another public action it will send a new message $m'=\langle a'_p, s'_p, idx_1, ..., idx_n \rangle$. The private state indexes in $m'$ for all agents, except $i$ are identical to the indexes in $m$.

An agent $j$ that receives both messages $m$ and $m'$ can compare them and identify that only the public state, and the state index for $i$, have changed. Thus, agent $j$ can know that message $m'$ contains a state that was generated in the same search path after $m$. This may reveal to agent $j$ a private dependency of agent $i$ between $a_p$ and $a'_p$.

To limit the private dependencies that are revealed in \mafs, each agent must initially decide on which dependencies it does not reveal (Line~\ref{line:initDependenciesNotAllowed} in Algorithm~\ref{alg:MAFS}). Then, whenever the situation above occurs, if there is a dependency for agent $i$ between $a_p$ and $a'_p$ that it does not wish to disclose, it will not broadcast $m'$ (Line~\ref{line:testDependency} in Algorithm~\ref{alg:MAFS}). Agent $i$ will not continue searching from the state that $m'$ represents, because even if we find a plan that uses that state, the final plan will leak the dependency we did not wish to disclose.
This modification --- limiting the broadcasted states --- allows \mafs to perform a joint search disclosing only some dependencies.

One can also envision an iterative method, where initially, all agents attempt solving without revealing any dependencies. If the agents are unable to produce a solution, each agent reveals one dependency, and so forth.
To avoid repeated expansions, each agent can save the unbroadcasted states mapped by the undisclosed dependency which was the reason that the state was not broadcasted.  When revealing a new dependency, the agent can publish all  the states that were not previously published due to the newly unrevealed dependency. We leave further exploration of this idea to future research.

\begin{table*}[ht]
\centering
\resizebox{0.9\textwidth}{!}{
\begin{tabular}{|l|c||c|c|c|c|c|c|c|c||c|c|c|c|c|c|c|c|}
\hline
\multicolumn{1}{|c|}{\multirow{3}{*}{\textbf{Domain}}} & \multirow{3}{*}{\textbf{\#dep}} & \multicolumn{8}{c||}{\textbf{Joint Projection}}                                                                                           & \multicolumn{8}{c|}{\textbf{MAFS}}                                                                                                        \\ 
\multicolumn{1}{|c|}{}                                 &                                 & \multicolumn{2}{c|}{\textbf{$m_1$}} & \multicolumn{2}{c|}{\textbf{$m_2$}} & \multicolumn{2}{c|}{\textbf{$m_3$}} & \multicolumn{2}{c||}{\textbf{$m_4$}} & \multicolumn{2}{c|}{\textbf{$m_1$}} & \multicolumn{2}{c|}{\textbf{$m_2$}} & \multicolumn{2}{c|}{\textbf{$m_3$}} & \multicolumn{2}{c|}{\textbf{$m_4$}} \\ 
\multicolumn{1}{|c|}{}                                 &                                 & \textbf{$\#C$}  & \textbf{$M_D\%$}  & \textbf{$\#C$}  & \textbf{$M_D\%$}  & \textbf{$\#C$}  & \textbf{$M_D\%$}  & \textbf{$\#C$}  & \textbf{$M_D\%$}  & \textbf{$\#C$}  & \textbf{$M_D\%$}  & \textbf{$\#C$}  & \textbf{$M_D\%$}  & \textbf{$\#C$}  & \textbf{$M_D\%$}  & \textbf{$\#C$}  & \textbf{$M_D\%$}  \\ \hline
BlocksWorld                                            & 1225                            & 19            & \textbf{53\%}    & 19            & \textbf{53\%}    & 19            & 62\%             & 19            & 62\%             & 14            & \textbf{69\%}    & 14            & \textbf{69\%}    & 14            & \textbf{69\%}    & 14            & \textbf{69\%}    \\ \hline
Depot                                                  & 6753                            & 20            & \textbf{35\%}    & 20            & 45\%             & 20            & 40\%             & 20            & 40\%             & 20            & \textbf{70\%}    & 20            & \textbf{70\%}    & 20            & 75\%             & 20            & 75\%             \\ \hline
Driverlog                                              & 5060                            & 19            & \textbf{8\%}     & 18            & 13\%             & 19            & 38\%             & 18            & 32\%             & 18            & \textbf{17\%}    & 18            & \textbf{17\%}    & 18            & 50\%             & 18            & 50\%             \\ \hline
Elevators                                              & 6245                            & 20            & 20\%             & 20            & 20\%             & 20            & \textbf{5\%}     & 20            & 60\%             & 20            & 51\%             & 20            & 51\%             & 20            & \textbf{13\%}    & 20            & 60\%             \\ \hline
Logistics                                              & 150                             & 21            & \textbf{10\%}    & 21            & 15\%             & 21            & \textbf{10\%}    & 21            & 15\%             & 20            & \textbf{10\%}    & 20            & 15\%             & 20            & \textbf{10\%}    & 20            & 15\%             \\ \hline
Rovers                                                 & 42155                           & 20            & \textbf{5\%}     & 20            & \textbf{5\%}     & 19            & 11\%             & 17            & 17\%             & 20            & \textbf{5\%}     & 20            & \textbf{5\%}     & 19            & 11\%             & 17            & 17\%             \\ \hline
ZenoTravel                                             & 1320                            & 17            & 25\%             & 16            & 4\%              & 18            & 55\%             & 18            & \textbf{5\%}     & 18            & 85\%             & 17            & 4\%              & 17            & 4\%              & 18            & \textbf{5\%}     \\ \hline
\end{tabular}
}
\caption{
Coverage and disclosed dependencies for the projection method and \mafs.
\#dep is the maximal amount of dependencies for a problem in the domain.
$\#C$ is the coverage for each heuristic $m_1,...,m_4$. 
$M_D\%$ is the percentage of the minimal amount of disclosed dependencies that allows solving the most problems. 
The heuristic with the minimal $M_D\%$ for each domain is in bold.}
\label{table:CoverageTable}
\end{table*}

\begin{figure*}[t!]
\centering
\begin{subfigure}[b]{0.3\textwidth}
\centering
  \includegraphics[width=1\linewidth]{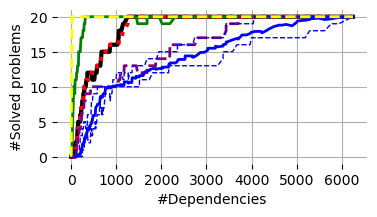}
  \caption{Elevators, max dep = 6245}
  \label{fig:Elevators}
\end{subfigure}\hspace{1em}
\begin{subfigure}[b]{0.3\textwidth}
\centering
  \includegraphics[width=1\linewidth]{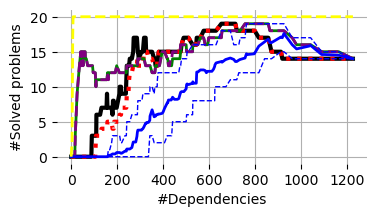}
  \caption{Blocksworld, max dep = 1225}
  \label{fig:Blocksworld}
\end{subfigure}\hspace{1em}
\begin{subfigure}[b]{0.3\textwidth}
\centering
  \includegraphics[width=1\linewidth]{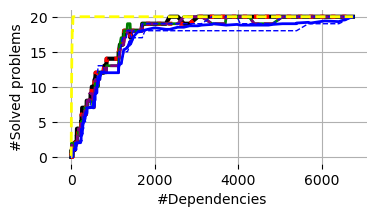}
  \caption{Depot, max dep = 6753}
  \label{fig:Depot}
\end{subfigure}\hspace{1em}
\begin{subfigure}[b]{0.3\textwidth}
\centering
  \includegraphics[width=1\linewidth]{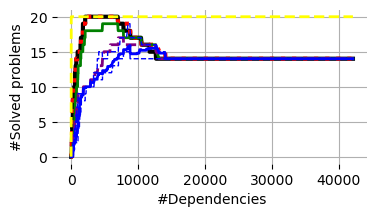}
  \caption{Rovers, max dep = 42155}
  \label{fig:Rovers}
\end{subfigure}\hspace{1em}
\begin{subfigure}[b]{0.3\textwidth}
\centering
  \includegraphics[width=1\linewidth]{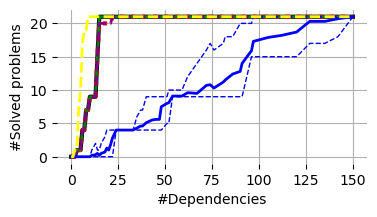}
  \caption{Logistics, max dep = 150}
  \label{fig:Logistics}
\end{subfigure}\hspace{1em}
\begin{subfigure}[b]{0.3\textwidth}
\centering
  \includegraphics[width=1\linewidth]{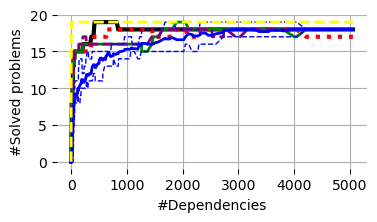}
  \caption{Driverlog, max dep = 5060}
  \label{fig:Driverlog}
\end{subfigure}\hspace{1em}
\begin{subfigure}[b]{0.3\textwidth}
\centering
  \includegraphics[width=1\linewidth]{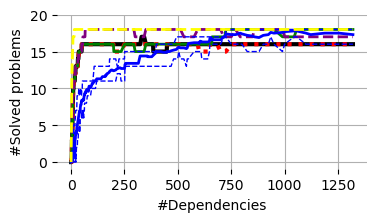}
  \caption{ZenoTravel, max dep = 1320}
  \label{fig:ZenoTravel}
\end{subfigure}\hspace{1em}
\begin{subfigure}[b]{0.5\textwidth}
\centering
  \includegraphics[scale=.7]{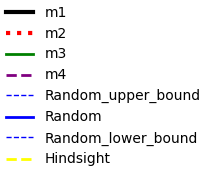}
  \caption{Legend.}
  \label{fig:Legend}
\end{subfigure}\hspace{1em}
\caption{\# of solved problems for each amount of published dependencies in the Joint projection method.
}
\label{fig:Coverage}
\end{figure*}

\begin{figure*}[t!]
\centering
\begin{subfigure}[b]{0.3\textwidth}
\centering
  \includegraphics[width=1\linewidth]{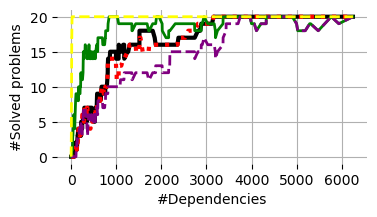}
  \caption{Elevators, max dep = 6245}
  \label{fig:ElevatorsMAFS}
\end{subfigure}\hspace{1em}
\begin{subfigure}[b]{0.3\textwidth}
\centering
  \includegraphics[width=1\linewidth]{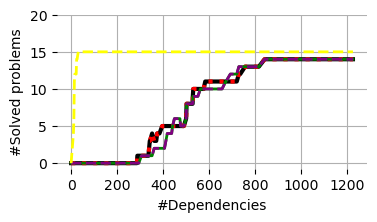}
  \caption{Blocksworld, max dep = 1225}
  \label{fig:BlocksworldMAFS}
\end{subfigure}\hspace{1em}
\begin{subfigure}[b]{0.3\textwidth}
\centering
  \includegraphics[width=1\linewidth]{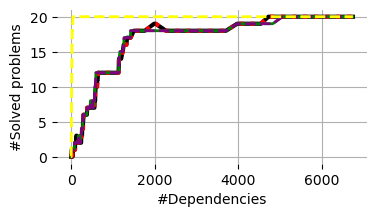}
  \caption{Depot, max dep = 6753}
  \label{fig:DepotMAFS}
\end{subfigure}\hspace{1em}
\begin{subfigure}[b]{0.3\textwidth}
\centering
  \includegraphics[width=1\linewidth]{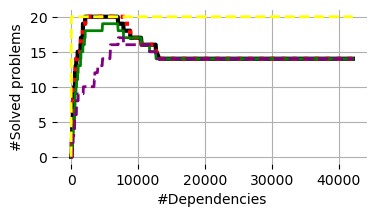}
  \caption{Rovers, max dep = 42155}
  \label{fig:RoversMAFS}
\end{subfigure}\hspace{1em}
\begin{subfigure}[b]{0.3\textwidth}
\centering
  \includegraphics[width=1\linewidth]{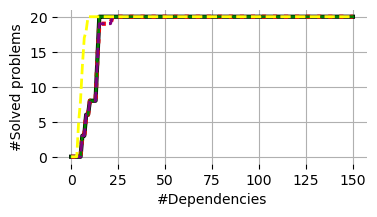}
  \caption{Logistics, max dep = 150}
  \label{fig:LogisticsMAFS}
\end{subfigure}\hspace{1em}
\begin{subfigure}[b]{0.3\textwidth}
\centering
  \includegraphics[width=1\linewidth]{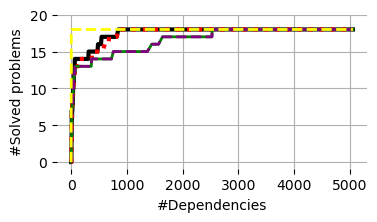}
  \caption{Driverlog, max dep = 5060}
  \label{fig:DriverlogMAFS}
\end{subfigure}\hspace{1em}
\begin{subfigure}[b]{0.3\textwidth}
\centering
  \includegraphics[width=1\linewidth]{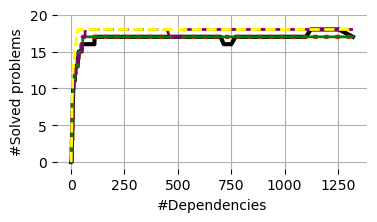}
  \caption{ZenoTravel, max dep = 1320}
  \label{fig:ZenoTravelMAFS}
\end{subfigure}\hspace{1em}
\begin{subfigure}[b]{0.5\textwidth}
\centering
  \includegraphics[scale=.7]{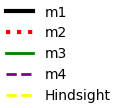}
  \caption{Legend.}
  \label{fig:LegendMAFS}
\end{subfigure}\hspace{1em}
\caption{\# of solved problems for each amount of published dependencies in the \mafs method.
}
\label{fig:CoverageMAFS}
\end{figure*}

\begin{figure*}[t!]
\centering
\begin{subfigure}[b]{0.3\textwidth}
\centering
  \includegraphics[width=1\linewidth]{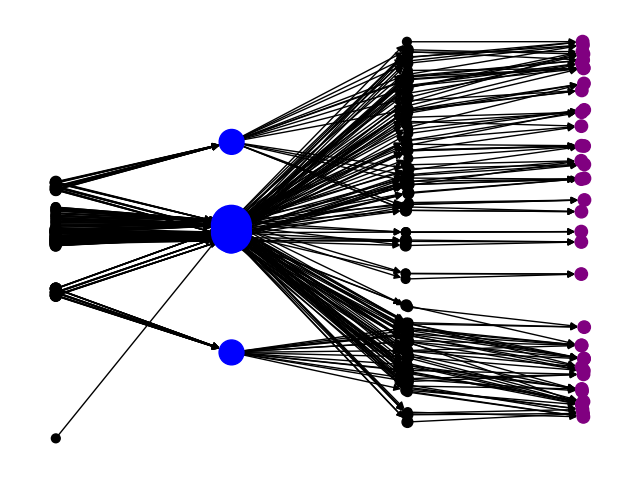}
  \caption{Elevators}
  \label{fig:DepGraphElevators}
\end{subfigure}\hspace{1em}
\begin{subfigure}[b]{0.3\textwidth}
\centering
  \includegraphics[width=1\linewidth]{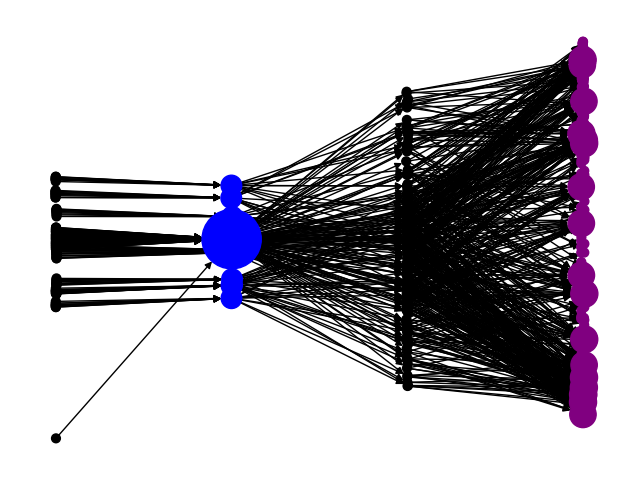}
  \caption{BlocksWorld}
  \label{fig:DepGraphBlocksWorld}
\end{subfigure}\hspace{1em}
\begin{subfigure}[b]{0.3\textwidth}
\centering
  \includegraphics[width=1\linewidth]{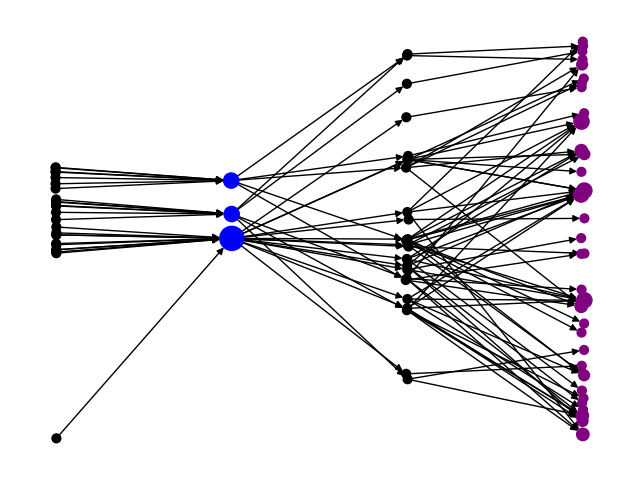}
  \caption{Depot}
  \label{fig:DepGraphDepot}
\end{subfigure}\hspace{1em}
\begin{subfigure}[b]{0.3\textwidth}
\centering
  \includegraphics[width=1\linewidth]{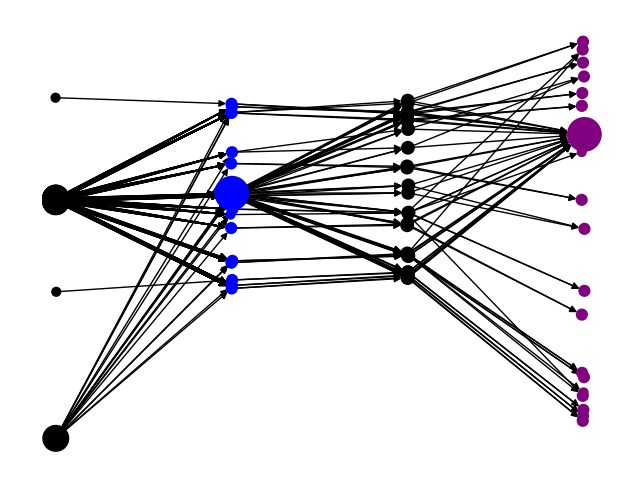}
  \caption{Rovers}
  \label{fig:DepGraphRovers}
\end{subfigure}\hspace{1em}
\begin{subfigure}[b]{0.3\textwidth}
\centering
  \includegraphics[width=1\linewidth]{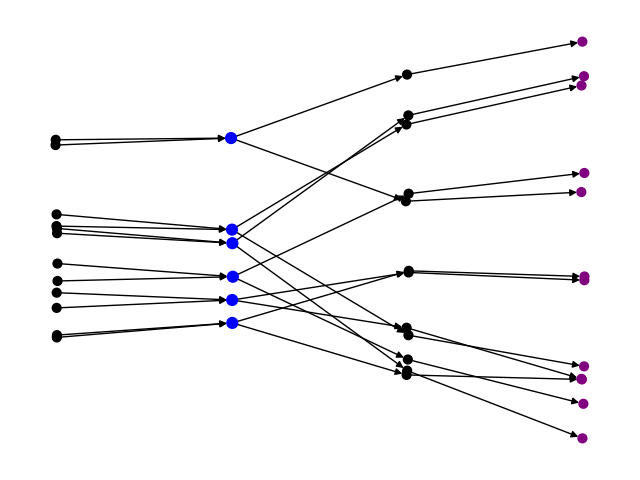}
  \caption{Logistics}
  \label{fig:DepGraphLogistics}
\end{subfigure}\hspace{1em}
\begin{subfigure}[b]{0.3\textwidth}
\centering
  \includegraphics[width=1\linewidth]{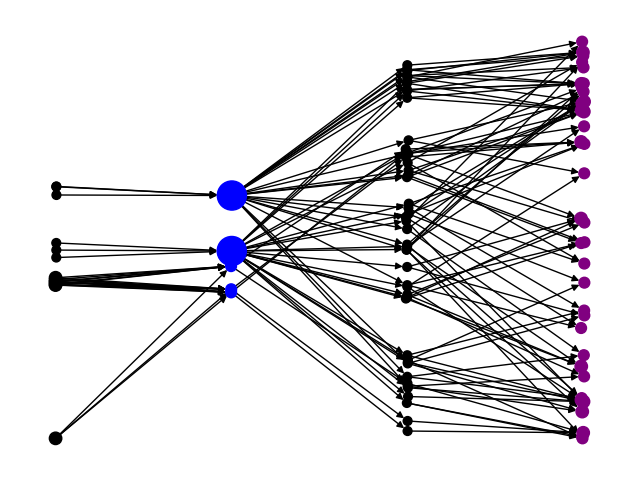}
  \caption{Driverlog}
  \label{fig:DepGraphDriverlog}
\end{subfigure}\hspace{1em}
\begin{subfigure}[b]{0.3\textwidth}
\centering
  \includegraphics[width=1\linewidth]{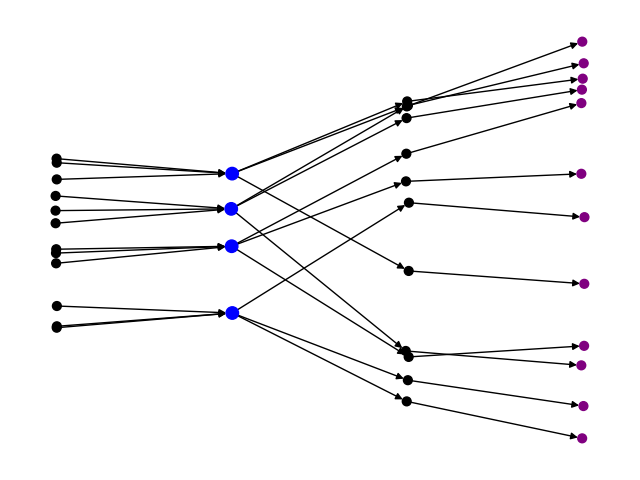}
  \caption{ZenoTravel}
  \label{fig:DepGraphZenoTravel}
\end{subfigure}\hspace{1em}
\caption{Single agent dependency graphs for all domains. All of the dependencies graphs are drawn for the easiest problem of the domain. Each dependencies graph has the same layers as described previously. The size of the nodes in the first and second layers, is their \emph{out-degree}, and the size of the nodes in the third and fourth layers is their \emph{in-degree}. There is also a special action that is a dummy-init-action in each domain, that action is at the bottom of the first actions layer. The dummy-init-action does not appear in Logistics and ZenoTravel, because it does not have any dependency with another public action in these domains.}
\label{fig:DependenciesGraphs}
\end{figure*}

\section{Empirical Analysis}

We now evaluate our methods using standard benchmarks \cite{vstolba2015competition}. For each problem in each domain, we ran the projection-based solver \cite{maliah2016projections} and the revised \mafs, both with a growing number of revealed dependencies. We denote the first as ``joint projection'' and the latter as ``\mafs''. The methods were implemented in C\# and run on a Windows machine with an i7-8700 3.20GHz CPU with 6 cores, and 16 GB of RAM.

In every experiment, we started without disclosing any private dependencies and run the chosen solver, recording whether it was able to solve the problem and, if so, the resulting plan makespan cost. 
We repeat this process, disclosing new private dependencies, running the solver, and recording the same statistics. Then, we compute the average solution cost and the number of solved problems under a timeout of 5 minutes. This value is referred to as \emph{coverage}.

\paragraph{Coverage Results}

Table~\ref{table:CoverageTable} shows the coverage results for the different solvers using the different heuristics $m_1$, 
$m_2$, $m_3$, and $m_4$. 
As can be seen, $m_3$, maximizing the number of public actions that can be executed, performs the best for Elevators, for both joint projection and \mafs solvers by an order of magnitude.
In BlocksWorld with the joint projection solver, $m_3$ and $m_4$ perform worse than $m_1$ and $m_2$, but it is only because of a single problem  that was solved faster by $m_1$ and $m_2$. The rest of the problems required less dependencies by $m_3$ and $m_4$, as can be seen in figure~\ref{fig:Blocksworld}.
In Driverlog, $m_3$ performs worse than $m_1$.
In this domain, 16 problems were solved with exactly the same amount of dependencies for all heuristics. The last 3 problems required more dependencies in $m_3$ than in $m_1$.
In Logistics and ZenoTravel, all methods performed the same (except of one difficult problem).
In Rovers, $m_4$ performs the worst, and all other methods perform almost the same (again, except of one problem).
In general, \mafs is slightly worse than the projection, and requires more dependencies in order to solve the problems.

Figure~\ref{fig:Coverage} shows the number of problems that were solved on each domain given the number of revealed dependencies by each agent in the projection-based solver. 
For each problem, we also computed a ``Hindsight value'', which is the number of private dependencies used by the plan that was generated when all dependencies are known. 
The hindsight is a baseline for comparison, indicating how many private dependencies are enough to find a solution. The hindsight may not be optimal, and it may be possible to provide a solution with fewer dependencies. We also report a method that randomly decides which dependency to disclose, averaged over 10 runs.

\begin{table*}[ht]
\centering
\footnotesize
\begin{tabular}{|c|c|c|c|c|c|c||c|c|c|c|c|}
\hline
\multirow{2}{*}{\textbf{Domain}} & \multirow{2}{*}{\textbf{M}} & \multicolumn{5}{c||}{\textbf{Joint projection}}                                       & \multicolumn{5}{c|}{\textbf{MAFS}}                                                    \\ \cline{3-12} 
                                 &                             & \textbf{Min} & \textbf{Max} & \textbf{Min. dep} & \textbf{Max. dep} & \textbf{Imp.}    & \textbf{Min} & \textbf{Max} & \textbf{Min. dep} & \textbf{Max. dep} & \textbf{Imp.}   \\ \hline
\multirow{4}{*}{BlocksWorld}     & m1                          & 48.95        & 111.75       & 92.10             & 55.15             & 45.39\%          & 29.33        & 30.93        & 30.07             & 30.20             & \textbf{1.93\%} \\ \cline{2-12} 
                                 & m2                          & 51.15        & 95.65        & 79.50             & 55.15             & \textbf{36.25\%} & 29.33        & 30.93        & 30.07             & 30.20             & \textbf{1.93\%} \\ \cline{2-12} 
                                 & m3                          & 47.35        & 99.65        & 83.20             & 50.10             & 43.05\%          & 28.57        & 31.64        & 29.57             & 30.64             & 2.98\%          \\ \cline{2-12} 
                                 & m4                          & 47.65        & 97.95        & 81.60             & 50.10             & 42.61\%          & 28.57        & 31.64        & 29.57             & 30.64             & 2.98\%          \\ \hline
\multirow{4}{*}{Depot}           & m1                          & 21.35        & 29.50        & 28.75             & 21.55             & 21.61\%          & 21.15        & 21.70        & 21.25             & 21.55             & 1.13\%          \\ \cline{2-12} 
                                 & m2                          & 21.35        & 29.45        & 28.40             & 21.55             & 19.73\%          & 21.15        & 21.65        & 21.20             & 21.55             & \textbf{0.71\%} \\ \cline{2-12} 
                                 & m3                          & 21.45        & 28.90        & 28.45             & 21.55             & \textbf{19.18\%} & 21.25        & 21.65        & 21.30             & 21.55             & \textbf{0.71\%} \\ \cline{2-12} 
                                 & m4                          & 21.45        & 29.05        & 28.60             & 21.55             & 20.26\%          & 21.25        & 21.65        & 21.30             & 21.55             & \textbf{0.71\%} \\ \hline
\multirow{4}{*}{Driverlog}       & m1                          & 28.89        & 63.63        & 61.42             & 29.32             & 32.82\%          & 21.11        & 23.56        & 22.39             & 22.67             & 6.48\%          \\ \cline{2-12} 
                                 & m2                          & 22.22        & 43.61        & 41.28             & 22.67             & 30.10\%          & 21.61        & 23.50        & 22.61             & 22.67             & 5.91\%          \\ \cline{2-12} 
                                 & m3                          & 39.21        & 58.63        & 57.00             & 39.47             & \textbf{23.98\%} & 21.17        & 23.39        & 22.17             & 22.67             & 5.91\%          \\ \cline{2-12} 
                                 & m4                          & 22.44        & 46.33        & 44.61             & 22.67             & 28.31\%          & 21.17        & 23.39        & 22.06             & 22.67             & \textbf{5.55\%} \\ \hline
\multirow{4}{*}{Elevators}       & m1                          & 33.50        & 53.10        & 37.45             & 46.90             & \textbf{10.67\%} & 37.25        & 61.75        & 39.90             & 55.70             & 6.88\%          \\ \cline{2-12} 
                                 & m2                          & 33.55        & 54.45        & 39.15             & 46.90             & 16.08\%          & 39.35        & 61.40        & 42.65             & 55.70             & 9.67\%          \\ \cline{2-12} 
                                 & m3                          & 33.00        & 51.25        & 38.35             & 46.90             & 13.38\%          & 33.60        & 58.90        & 36.10             & 55.70             & \textbf{6.36\%} \\ \cline{2-12} 
                                 & m4                          & 34.75        & 51.75        & 38.35             & 46.90             & 10.20\%          & 37.55        & 60.50        & 40.90             & 55.70             & 9.19\%          \\ \hline
\multirow{4}{*}{Logistics}       & m1                          & 25.05        & 31.29        & 27.24             & 29.24             & 6.40\%           & 25.15        & 31.95        & 27.85             & 29.35             & 8.04\%          \\ \cline{2-12} 
                                 & m2                          & 25.05        & 31.29        & 26.81             & 29.24             & \textbf{5.33\%}  & 25.15        & 31.95        & 27.35             & 29.35             & \textbf{6.75\%} \\ \cline{2-12} 
                                 & m3                          & 25.00        & 31.24        & 27.24             & 29.24             & 6.62\%           & 25.15        & 31.95        & 27.85             & 29.35             & 8.07\%          \\ \cline{2-12} 
                                 & m4                          & 25.00        & 31.24        & 26.81             & 29.24             & 5.55\%           & 25.15        & 31.95        & 27.35             & 29.35             & 6.79\%          \\ \hline
\multirow{4}{*}{Rovers}          & m1                          & 49.50        & 53.00        & 52.40             & 50.25             & 6.07\%           & 57.50        & 62.30        & 61.45             & 58.35             & 8.12\%          \\ \cline{2-12} 
                                 & m2                          & 50.25        & 51.50        & 51.50             & 50.25             & \textbf{3.28\%}  & 58.25        & 60.40        & 60.40             & 58.35             & \textbf{4.99\%} \\ \cline{2-12} 
                                 & m3                          & 51.74        & 68.42        & 67.58             & 52.84             & 22.68\%          & 60.26        & 83.05        & 82.00             & 62.37             & 26.83\%         \\ \cline{2-12} 
                                 & m4                          & 51.53        & 64.35        & 63.35             & 52.53             & 18.26\%          & 58.00        & 77.41        & 77.41             & 58.41             & 24.41\%         \\ \hline
\multirow{4}{*}{ZenoTravel}      & m1                          & 45.12        & 56.12        & 47.59             & 54.41             & 9.32\%           & 52.44        & 63.11        & 55.00             & 61.28             & 8.96\%          \\ \cline{2-12} 
                                 & m2                          & 37.31        & 50.63        & 39.81             & 48.81             & 9.81\%           & 44.06        & 56.53        & 46.65             & 54.59             & 9.39\%          \\ \cline{2-12} 
                                 & m3                          & 51.83        & 62.83        & 54.61             & 60.72             & \textbf{8.69\%}  & 45.24        & 56.71        & 47.94             & 54.59             & 9.06\%          \\ \cline{2-12} 
                                 & m4                          & 51.83        & 62.83        & 54.61             & 60.72             & \textbf{8.69\%}  & 52.33        & 63.39        & 55.11             & 61.28             & \textbf{8.69\%} \\ \hline \hline
\multirow{4}{*}{Average}         & m1                          & 36.05        & 56.91        & 49.56             & 40.97             & 18.90\%          & 34.85        & 42.19        & 36.84             & 39.87             & 5.93\%          \\ \cline{2-12} 
                                 & m2                          & 34.41        & 50.94        & 43.78             & 39.22             & \textbf{17.23\%} & 34.13        & 40.91        & 35.85             & 38.91             & \textbf{5.62\%} \\ \cline{2-12} 
                                 & m3                          & 38.51        & 57.27        & 50.92             & 42.98             & 19.66\%          & 33.61        & 43.90        & 38.13             & 39.55             & 8.56\%          \\ \cline{2-12} 
                                 & m4                          & 36.38        & 54.79        & 48.28             & 40.53             & 19.13\%          & 34.86        & 44.28        & 39.10             & 39.94             & 8.33\%          \\ \hline
\end{tabular}
\caption{Averaged plan makespan cost over all of solved problems for each domain. Min denotes the average minimal cost over all solved problems.
Max denotes the average maximal cost.
Min. dep denotes the cost when solved with the minimal amount of disclosed dependencies.
Max. dep denotes the cost when solved with the maximal amount of disclosed dependencies.
Imp. denotes improvement between the first time that the problem was solved to the best achieved solution.
The lowest improvement, and hence best initial plan, is in bold.}
\label{table:costTable}
\end{table*}

$m_3$, which prioritizes enabling additional public actions, performs the best in all domains. The rest of the methods vary in their performance. On Blocksworld, e.g., $m_4$, that prioritizes achieving additional public facts, is the best together with $m_3$, but on Elevators and Rovers $m_4$ performs the worst. On Elevators, arguably the domain with the highest amount of required collaboration, differences between methods are most pronounced. On Depot, random dependencies have identical performance as the heuristics, except for 2 problems, where the random selection did not identify needed dependencies.

Figure~\ref{fig:DependenciesGraphs} shows agent's dependency graphs  (Section~\ref{scn:RankPublishedDep}).
In BlocksWorld and Elevators, $m_3$ performs the best due to an artificial effect that is precondition to many actions, manifesting as a larger blue node in Figure~\ref{fig:DependenciesGraphs}. $m_1$ and $m_2$ reveal this artificial effect numerous times, instead of other useful effects, while $m_3$ reveals this effect once, and publishes other effects to achieve more actions.

$m_4$ performs very well on BlocksWorld but poorly on Elevators and Rovers. This is because BlocksWorld has some public facts that are easier to achieve, as they are achieved by many actions (denoted by larger purple nodes). $m_4$ avoids publishing additional ways to achieve these facts, and can hence allow achieving other useful public facts. In the other domains, most public facts have almost the same number of actions that can achieve them. $m_4$ cannot distinguish between these dependencies.
Hence, an analysis of the agent's dependency graph can reveal whether $m_3$ or $m_4$ can be more useful. $m_4$ performs better under diverse in-degree for the purple nodes, while $m_3$ performs better given a uniform in-degree.

On Depot all methods perform much worse than the hindsight, and very similar to the random solver, leaving much room for improvement.
Looking at the dependency graph (Figure~\ref{fig:DepGraphDepot}), for each public fact (purple node) there is a very small number of paths to it. Hence, practically every dependency seems equally important to all heuristics, yet many of them are not useful for achieving the goal.

An interesting phenomenon occur in, e.g., BlocksWorld and Rovers, where some problems are solved when not all dependencies are available, but cannot be solved when more dependencies are published. This is because the projection method may produce a public plan that cannot be extended into a complete plan, and we did not backtrack. Our methods were often able to publish dependencies that resulted in plans that could be extended. Later, additional dependencies confused the planner to choose plans that could not be extended.

Figure~\ref{fig:CoverageMAFS} shows the number of problems that were solved on each domain given the number of revealed dependencies by each agent using \mafs.  As in the projection-based solver, $m_3$ seems to be the best here in all the domains.
In \mafs, for Elevators the difference between the heuristic methods is most pronounced.
\mafs performed worse than the projection method in our experiments, but in BlocksWorld the results are very poor, as \mafs timed out on many problems. This happened because the projection method was better at identifying plans with very little agent collaboration, while \mafs tended to identify plans here with significantly more collaboration.

\paragraph{Plan Cost Results}

Finally, we consider the makespan cost of the generated public plan. Dependencies affect not just solvability but also cost. Even when a plan could be obtained without revealing any dependency, better plans may be obtained by collaboration, revealing some dependencies. Thus, agents may  trade off some privacy for increased efficiency.

Table~\ref{table:costTable} shows the results of the following averaged factors: (1) Min. cost, (2) Max. cost, (3) Min. dep. cost --- the solution cost for a plan found with the least amount of revealed dependencies, 
and (4) Max. dep. cost -- the solution cost for a plan that was found with the maximal amount of dependencies revealed.
The ``Improvement'' column  shows  the percentage of the minimal cost out of the Min. dep. cost $(\frac{Min. \, dep. \, cost - Min. \, cost}{Min. \, dep. \, cost}*100\%)$. That is, how much cost can be reduced by revealing more dependencies. 
On average, the improvement is only about 20\% for the different methods in the joint projection planner, and only 10\% in \mafs. This means that the solution obtained with the least amount of revealed dependencies was almost as good as the best solution that was found when we revealed additional dependencies.

However, the impact of sharing private dependencies on solution cost varies significantly per domain. 
For example, in the joint projection solver, in BlocksWorld, the improvement is about 40\%. This is because every problem can be solved by a single agent doing all the work. However, if the agents work jointly, they distribute the actions to provide a better makespan. 
Also, in most cases, there was almost no difference between the different heuristics ($m_{1-4}$) with respect to the solution cost. This shows that our heuristics are better geared towards solvability, not makespan. We leave heuristics oriented towards reducing costs to future research.

The improvement made by \mafs is less than the improvement made by the joint projection solver. 
This is because \mafs prefers plans with more collaboration.

\section{Conclusion and Future Work}

In this work we suggest methods for publishing only a part of the private dependencies between public actions of agents. We provide experimental results on both the joint projection method that uses all private dependencies to compute a public plan, and \mafs, modified to broadcast states only if they do not reveal a forbidden dependency. We show that in many cases a public plan can be computed with only a small portion of the dependencies, and that our heuristic methods rapidly find a good subset to share for both the joint projection method and \mafs.
We provide experiments over standard benchmark domains, comparing the coverage of our methods, as well as the cost of the found plans. 

Future work should analyze the privacy leakage, with respect to the revealed transition systems \cite{StolbaFK19} in the context of using the shared dependencies in current solvers.
It is also worthwhile to assign importance values to different dependencies, where each dependency has an intrinsic value that needs to be considered.

\section*{Acknowledgements}
This work is partially funded by ISF grant \# 210/17 to Roni Stern and by ISF grant \# 1210/18 to Guy Shani.

\bibliographystyle{named}
\bibliography{main}

\end{document}